\documentclass[twocolumn,preprintnumbers,amsmath,amssymb,prl]{revtex4}
\usepackage{graphicx}
\usepackage{dcolumn}
\usepackage{bm}

\begin{document}
\title{Zipping and unzipping of nanoscale carbon structures}
\author{Julia Berashevich and Tapash Chakraborty}
\affiliation{Department of Physics and Astronomy, 
University of Manitoba, Winnipeg, Canada, R3T 2N2}
\begin{abstract}
We demonstrate theoretically that hydrogenation and annealing 
applied to nanoscale carbon structures play a crucial role in determining
the final shape of the system. In particular,
graphene flakes characterized 
by the linear and non-hydrogenated zigzag or armchair edges
have high propensity to merge into a 
bigger flake or a nanotube 
(the formation of a single carbon-carbon bond
lowers the total energy of the system by up to 6.22 eV). 
Conversely, the line of the $sp^2$ carbon bonds 
(common for pure carbon structures such as graphene or a carbon nanotube)
converted into the $sp^3$ type by hydrogenation shows an ability to 
disassemble the original structure by cutting it along the line 
of the modified bonds. These structural transformations provide us with
an understanding of the behavior of 
mobile carbon structures in solution and a distinct scenario of how to preserve 
the original structure which would be a crucial issue 
for their application in carbon-based electronics.
\end{abstract}

\maketitle
\section{Introduction}
The unique electronic properties of graphene \cite{novos,review}, in particular the
high mobility of its charge carriers
make this recently discovered material an
excellent candidate for semiconductor electronics \cite{berger,chen}.
The band gap of the finite size graphene is 
defined by the $\pi$ states induced by the unsaturated dangling bonds at the edges,
and therefore, the electronic properties of the finite size graphene structures
strictly depend on the size and geometry of the 
flakes and also on their atomic edge structure.
Since these $\pi$ states are spin-polarized,
the spin polarization of graphene will depend on the
atomic edge structure as well. 
It was revealed that under certain conditions
(for example in an applied electric field \cite{son} 
or doping of the zigzag edges \cite{ber1})
the spin-up and spin-down states can be
separated between two zigzag edges 
such that the half-metallicity is achieved. Additionally, 
if an imbalance in concentration of the 
spin-up and spin-down states are to be introduced 
through edge modification, it would result in non-zero magnetization \cite{yazyev}.
It is then expected that the size of the graphene flakes and the
atomic structure of the zigzag edges are the most crucial 
parameters for application of graphene in semiconductor electronics and 
spintronics. It is already known that the standard chemical methods 
used to fabricate the graphene flakes 
lead to the rough edges \cite{nov,lix,campos}. 
Therefore, for the rapidly developing field of carbon electronics,
the technology of "cutting" graphene into
nanoscale flakes of controllable shape with linear edges
is immensely important. Interestingly, there is also a growing demand  
for generation of large graphene sheets which have already found their applications in 
touch screens and solar cells \cite{techn_rev}.

Development of the technology for cutting the carbon structures 
has been under consideration since the last decade
and began with the discovery of carbon nanotubes \cite{moon}.
Originally it was observed that cutting and unzipping of nanotubes 
can be performed in highly acidic environment
where hydrogen ions supposedly break the carbon-carbon bonds, while 
further annealing restores some of the tubular structures \cite{moon} and
restoration would be more sufficient at higher temperatures \cite{tang}.
It was also observed that annealing of nanotubes can lead
to coalescence of the nearest located tubes into
the one of a larger diameter \cite{nikolaev}.
Since then, several technologies have been developed to unzip nanotubes \cite{terrones}
but the most popular ones are chemical methods which
use oxidizing agents to break the carbon-carbon bonds \cite{kosynkin},
and catalytic hydrogenation using the metal nanoparticles as scissors
\cite{cluster,cluster1}. 
Therefore, the original idea \cite{moon} 
that nanotubes can be disintegrated or integrated due to the interaction 
with the atomic hydrogen lost its appeal with the discovery of these 
new chemical methods. Of late, it has received a fresh scrutiny
\cite{yang} after the discovery of graphene. These authors have used 
hydrogenation of graphene edges to control the nanoribbon width and 
the edge quality within the plasma etching procedure.

Therefore, in this work we wish to refocus the 
attention on the method of manipulating the 
carbon structure by their interaction with atomic hydrogen,
by providing insights on
the phenomena of merging and cutting processes 
in carbon structures with the help of the atomic
hydrogen. We have simulated the transformation (merging and unzipping)
of the carbon system such as graphene flakes and carbon nanotubes
in vacuum with the help of quantum chemistry methods. 
The calculations have been performed 
using the spin-polarized density 
functional theory with the hybrid functional 
UB3LYP/6-31G available in the Jaguar 7.5 program \cite{jaguar}.
For interaction and bonding of the carbon atoms with hydrogen atoms
the B3LYP exchange-correlation functional has been found to offer 
sufficient precision level for the geometry parameters and energetics \cite{lozyn} 
thus providing a good agreement with the results obtained via a
more accurate level such as the
{\it ab initio} M\o ller-Plesset second-order (MP2) method.
The 6-31G basis set was found to be adequate to describe the 
alteration of the total energies and lattice parameters in the carbon systems.
The initial structures of the graphene flakes have been generated using the 
standard parameters of graphene honeycomb lattice 
(the carbon-carbon bond length was 1.42 \AA). 
During the geometry optimization (UB3LYP/6-31G) the 
systems were allowed to relax in all directions thereby forming the 
shape characterized by the lower total energy. 
For all optimization procedures, the transition of the structure from an initial 
state to the optimized geometry (with corresponding structural changes of 
the graphene lattice) during the relaxation procedure occurs without any 
energetical barrier.
To perform the hydrogenation of the graphene flakes the 
hydrogen atoms have been attached to the required 
carbon atoms and their positions have been further optimized.

\begin{figure}
\includegraphics[scale=0.48]{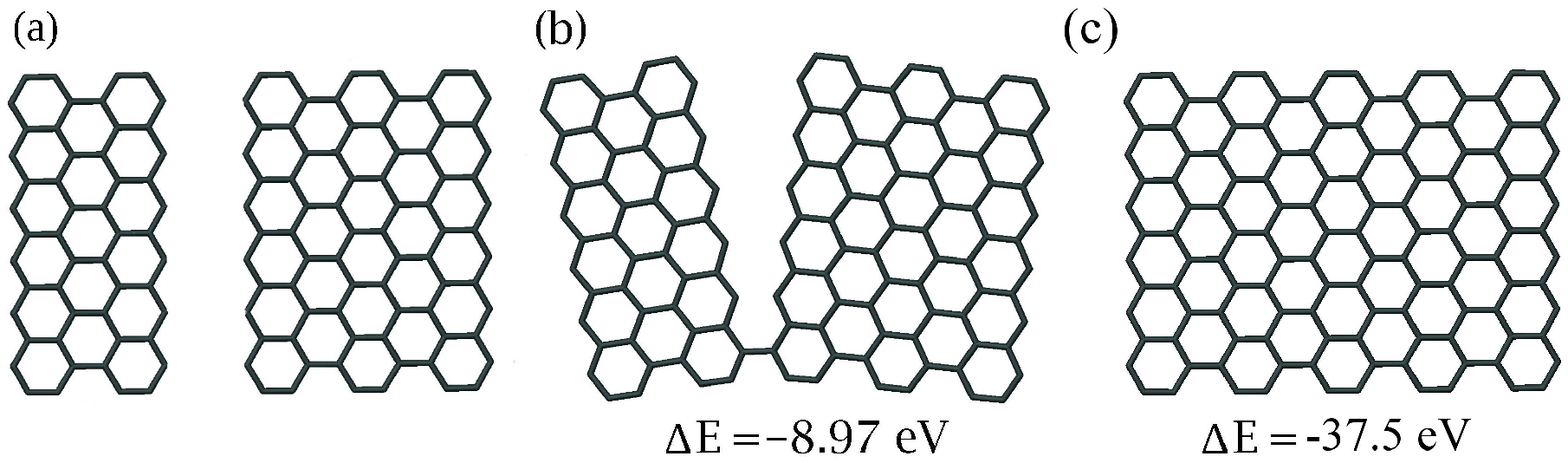}
\includegraphics[scale=0.42]{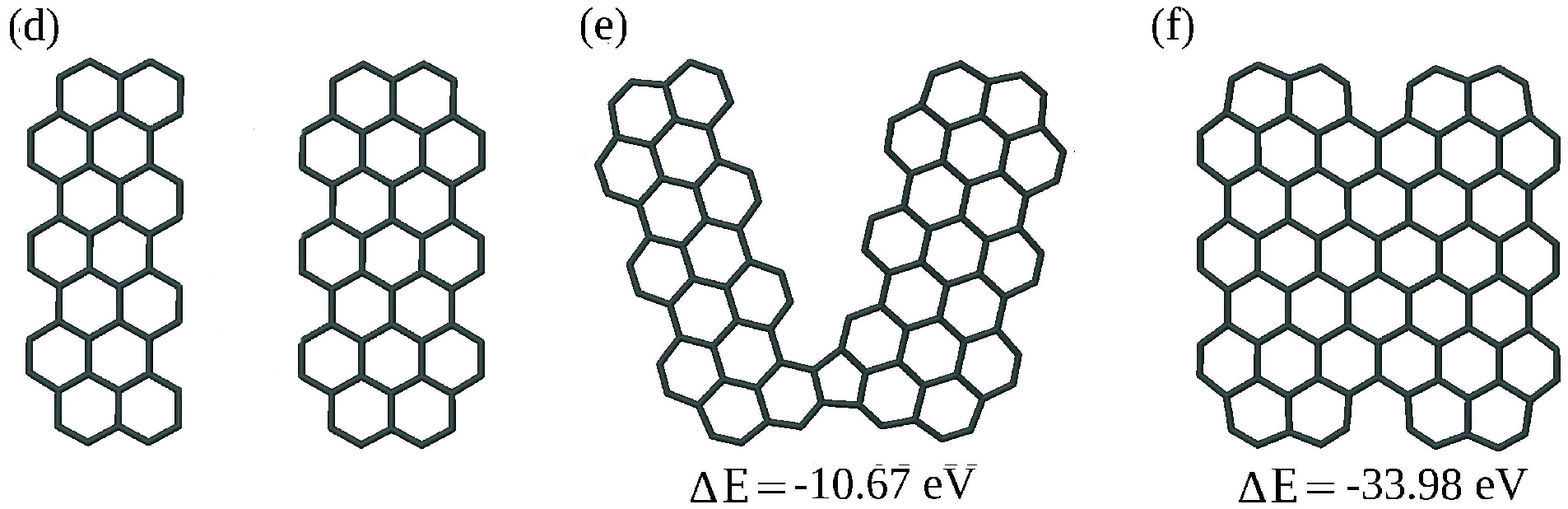}
\caption{Integration of two isolated flakes 
characterized by the sharp and linear edges 
(top panel presents the case when the flakes are stacked by the zigzag 
edges, while the bottom panel is for flakes stacked by the armchair edges)
into one flake of larger size:
a) and d) two initially separated flakes; b) and e) 
formation of the first carbon-carbon bond; 
c) and f) the merged systems.}
\label{fig:fig1}
\end{figure}

\section{Zipping process}
Recent developments in 
the technology of fabrication of carbon systems have demonstrated that the 
sharpness of zigzag edges in graphene can be achieved by Joule heating 
\cite{jia}, i.e., by evaporation of carbon atoms whose presence at the 
edges break the linearity of these edges. The same effect is also achieved 
by H$_{2}$-plasma etching \cite{yang}. These methods result in 
sharp and highly crystalline edges that, as we mentioned above, 
would promote the application of graphene 
in semiconductor electronics. In this respect we investigate the properties 
of graphene flakes of rectangular shape and found that
when their edges are linear and non-hydrogenated 
the interaction between the two flakes is so high 
that several flakes will have the tendency to merge into one big structure. 
This result is extremely important and requires considerable attention because it might have
valuable impact on the technology of fabrication of the 
touch screens and solar cells \cite{techn_rev} as a way
to create a graphene sheet of desirable size through flake integration.

For our investigations we took two graphene flakes with linear zigzag or armchair edges
and placed them side by side, as shown in 
Fig.~\ref{fig:fig1} (a) and (d). After relaxation of the system, 
two flakes were merged along the stacked edges, 
thereby generating one flake of bigger size. The flakes with the
linearly cut zigzag edges zip up along those zigzag edges 
(see (a-c) in Fig.~\ref{fig:fig1})
while in the other case the zipping occurred between the stacked armchair edges  
(see (d-f) in Fig.~\ref{fig:fig1}). We presented in Fig.~\ref{fig:fig1} three 
major steps of the merging process: the separated flakes, origin of the
zipping process, and the merged system.
Zipping begins by moving the flakes close to each other 
by one seam with the formation of the first carbon-carbon bond between 
the two flakes (see (b,e) in Fig.~\ref{fig:fig1}) and after that the rest of the 
system zip up (see (c,f) in Fig.~\ref{fig:fig1}). 
To show that zipping occurs because it is the energetically favored process,
we calculated the energy $\Delta E$ (see its value in Fig.~\ref{fig:fig1})
as the difference between the total energy of the original systems 
(Fig.~\ref{fig:fig1} (a) or (d)  
for the zigzag and armchair edges, respectively) and the
total energy of the structure taken in the current conformation.
The energy $|\Delta E|$ grows as the flakes move 
close to each other (Fig.~\ref{fig:fig1} (b) or (e)  
for the zigzag and armchair edges, respectively) 
due to the interaction between them, 
while subsequent formation of each carbon-carbon bond
increases that energy even further. 

Since formation of a large flake through 
zipping of two flakes of smaller size significantly lowers the total
energy of the system, it implies that this process 
should occur with any flakes having the following characteristics: 
non-hydrogenated linear edges and ability to migrate. 
The evolution of the total energy with zipping of the flakes is 
defined not only by the zipping mechanism itself but by other parameters such as 
the original flake conformation and the distance between the flakes. 
However, to gain insight on the merging phenomena
we need to estimate the energy released as each carbon-carbon bond is made. 
For this purpose we calculated the formation energy 
of the flake integration as $\Delta E_{z}=E_{(merged)}-(E_{1}+E_{2})$, where 
$E_{(merged)}$ is the total energy of the system after integration, while
$E_{1},E_{2}$ are the total energies of the initially separated first and second flakes, respectively. 
We found that in the case when zipping 
involves the zigzag edges (see (a-c) in Fig.~\ref{fig:fig1}), 
the formation energy $\Delta E_{z}$ was -43.99 eV,
while for zipping along the armchair edges
(see (d-f) in Fig.~\ref{fig:fig1}) $\Delta E_{z}$=-51.20 eV.
However, because the 
conformation of the graphene structure is different for the initial and final 
steps, the formation energy $\Delta E_{z}$ contains two major components and 
one of which is the energy released with the generation of each carbon-carbon 
bond while the second one is the reorganization energy of the lattice, i.e., the
energy required to reorganize the graphene lattice with a transition of the 
system from the initial to final states.
Therefore, to estimate the energy released by formation
of the single carbon-carbon bond,
we need to substitute the reorganization energy from the total 
formation energy.
We calculated the reorganization energy 
to be -6.63 eV for merging of the flakes through zipping of the zigzag edges and
-5.78 eV for flakes merging through zipping of the armchair edges.
As a result we found
that the formation energy of a single bond made between the zigzag 
edges is $\sim$ -6.22 eV. This value is close to that of the dissociation energy 
of the double carbon-carbon bonds which is $\sim$6.4 eV \cite{tut}.
For the armchair edges we obtained the 
formation energy of a single bond to be $\sim$ -5.92 eV, 
but this value can not be very accurate
because in the case of zipping of the armchair edges 
the situation is more complicated
as the armchair edges can not zip up completely 
without introducing some defected sites at the end of the line 
along which two flakes join (see Fig.~\ref{fig:fig1} (f))

\begin{figure}
\includegraphics[scale=0.45]{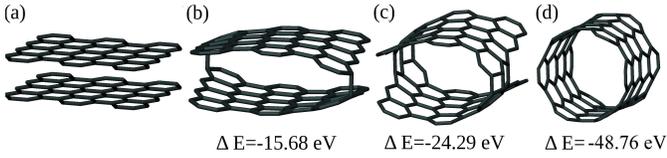}
\caption{Formation of a carbon nanotube from two graphene flakes
placed on top of each other
with linearly cut and non-hydrogenated zigzag edges: a) two separated flakes, b) 
origin of the zipping process, c) formation of the carbon nanotube.}
\label{fig:fig2}
\end{figure}

As it is impossible to merge the armchair edges without 
introducing the defected sites and because 
the zigzag edges seems to have a higher tendency 
to zip up than the armchair edges,
the zipping along the zigzag edges
should receive careful consideration for fabrication of
large graphene films required for 
touch screens and solar cells technology \cite{techn_rev}.
For these applications, the effect of the flake size (finite size effects) 
should be an important concern. Because the lattice reorganization
within the zipping process occurs mostly at the zipping edges 
the contribution of the reorganization energy to $\Delta E_z$ 
will gradually decrease with increasing length of the flake along the zipping edges.
A similar conclusion was reached for the second energy component (linked 
with the formation of the carbon-carbon bonds)
contributing to $\Delta E_z$, i.e. 
$\Delta E_z$ will be reduced with an increase of the number of 
carbon-carbon bonds that are to be formed. Therefore, an increase 
in flake size, namely the flake length along the axis of zipping, 
would favor an increase of energy $\Delta E_z$ released in the process.
Growth of the flake size in the other direction
will not drastically modify the energetics of the zipping process.
Moreover, the large formation energy of the carbon-carbon bond 
(-6.22 eV for the zigzag edges) suggests that the flake zipping 
would occur even at non-zero temperatures as this formation energy 
is much larger than the thermal energy ($k_BT$). The last conclusion 
is in fact, supported by the experimental observation 
of zipping of the damaged nanotubes \cite{nikolaev,mitra}.

Because generation of each carbon-carbon bond 
releases so much energy, 
placing the flakes with linearly cut zigzag edges 
on top of each other may lead to occurrence of the 
zipping process at both sides limited by the zigzag edges 
thereby creating a nanotube. 
We performed the relaxation of the system 
presented in Fig.~\ref{fig:fig2} (a) and 
achieved the rolling of the flakes into a nanotube 
(see Fig.~\ref{fig:fig2} (d)). 
At first, the graphene flakes 
slowly bend into a tubular form (see Fig.~\ref{fig:fig2} (c)) and 
with that bending the zigzag edges start to zip up (see Fig.~\ref{fig:fig2} (d)). 
We calculated the energy of nanotube formation to be
$\Delta E_{z}$=-47.86 eV. Because 
there are twice the carbon-carbon bonds 
to be made to generate a nanotube from two flakes, 
it was expected that the energy released due to the tube formation 
should be much larger than that for
integration of two flakes into one of bigger size.
To confirm that we took the same flakes as in Fig.~\ref{fig:fig2} (a) and 
placed them side by side, zipped them 
along their zigzag edges and calculated the formation energy
to be $\Delta E_{z}$=-36.58 eV.
As expected, we found that the generation of a nanotube
is much more favorable energetically in comparison to
flakes merging into a flake of larger size. 
For flakes presented in Fig.~\ref{fig:fig2} (a)
the energy difference between the two final shapes (a nanotube in comparison to a flake) 
achieves the magnitude of -13.86 eV and this difference will grow 
with elongation of the nanotube structure.
The reorganization energy in the case of the nanotube formation is 
quite small (-1.21 eV) which gives us the formation energy 
of the carbon-carbon bond to be -5.83 eV. 
In contrast, for the flakes merging into a flake of larger size,
the reorganization energy remains significant (-10.66 eV) giving a
bond formation energy of -6.48 eV.

Therefore, if there are two flakes with non-hydrogenated linear
zigzag edges, depending on certain conditions they 
can form either a flake of larger size or a carbon nanotube. Obviously,
the main condition controlling the final shape 
would be the mobility of the flakes and 
their original position relative to each other. 
Therefore, in a solution where the
carbon systems presumably have higher mobility, 
the energetically favored shape of the carbon structure 
after annealing is expected to be
a carbon nanotube. That is the reason for zipping of
partially unzipped carbon nanotubes back to the original nanotube
when its edges at the cutting front possess the dangling bonds \cite{tang}.
As formation of each carbon-carbon bond drastically
lowers the total energy of the system, the process involving more 
bond formation would be favored energetically.
Moreover, in a geometry of the carbon systems with 
less number of carbon atoms at the edges the total energy is lowered.
In fact, this observation offers an understanding 
of the experimental data on coalescence 
of the carbon nanotubes in solution \cite{nikolaev}. For example, 
we estimated that coalescence of two nanotubes 
(placed one behind other thereby forming a longer tube)
of the same diameter equal to that in Fig.~\ref{fig:fig2} (d)
lowers the total energy by 51.34 eV.

Another important understanding is that the absence of hydrogenation at the edges makes 
the graphene flakes chemically attractive to each other. If interaction of the 
graphene flake with the surrounding environment 
can be considered as weak \cite{ber} (because interaction 
with most of the adsorbates occurs via the van der Waals forces \cite{ort}), 
the interaction between the graphene flakes itself is quite significant  
and may lead to their integration. Therefore, if we are to prevent the 
integration of the carbon structures for their application in semiconductor electronics, 
they should be immobilized or being hydrogenated along the edges.

\section{Unzipping process}
For graphene hydrogenation, the edges are the ones to be first hydrogenated 
because the energy required to attach a single hydrogen atom 
to the edges is about 3.0 eV less than that for hydrogen attachment to the 
graphene surface. Hydrogenation of graphene is a
thermally activated process and is rarely found to
occur at room temperature \cite{ker,exp,exp1}. Therefore, 
it is possible to hydrogenate 
the graphene edges selectively from the rest of the structure by controlling the temperature.
For hydrogenation of the graphene surface, 
it appears that the conformation characterized by the lower total energy 
would be when the nearest-neighbor carbon atoms 
belonging to different sublattices (there are A and B sublattices in graphene \cite{review})
are hydrogenated from different sides of the graphene plane \cite{bouk}. 
In this case the chemisorption energy is characterized 
by the lowest value and if the surface of graphene is fully hydrogenated, the system will 
turn into graphane in its chair conformation \cite{sofo,galv}.

For partial hydrogenation of the graphene surface, 
if the first hydrogen atom is randomly attached to the graphene surface,
the second atomic hydrogen will be attached to the 
nearest-neighbor carbon atom as more energetically favorable position \cite{cas}.
Let us consider the process leading to the lowest state, i.e., to graphane in the chair conformation 
that occurs when hydrogenation is applied to both sides of the graphene plane.
In this case, attaching the second hydrogen atom to the nearest-neighbor carbon atom
with higher probability would occur at other side of the plane \cite{bouk}.
To continue the hydrogenation process 
we have to take into account the fact that the carbon atoms located closer to the already hydrogenated part
requires less energy to be hydrogenated as this energy increases with increasing 
distance from the already hydrogenated carbon atoms \cite{cas}.
Because the pair of hydrogen atoms attached to the nearest-neighbor carbon 
atoms weakens the neighboring bonds, under some conditions of hydrogenation 
the attachment of hydrogen atoms may occur linearly (when minor difference 
in magnitude of the binding energy tends to control the hydrogenation process).
As hydrogenation is a thermally activated process \cite{ker,exp,exp1},
such a control can be obtained at a lower temperature, while its substantial 
increase would actually lead to full hydrogenation of the graphene surface, 
i.e., to graphane-like structures. This effect of weakening of the neighboring 
bonds by an exposure to active adsorbates is used for longitudinal cutting 
of the carbon nanotubes by the oxidative process \cite{kosynkin,xu}
and has been predicted to be the case in hydrogenation of the 
nanotubes \cite{lu}.

\begin{figure}
\includegraphics[scale=0.45]{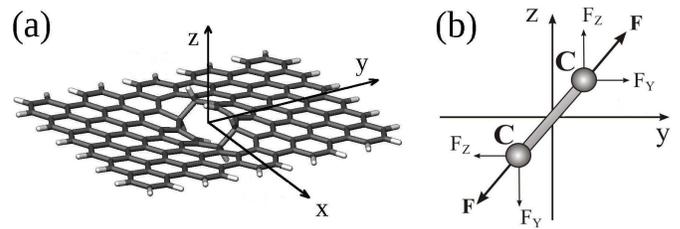}
\caption{(a) Breaking of the carbon-carbon bond as three pairs of hydrogen atoms 
attached to the carbon atoms line up; (b) distribution of the 
interatomic forces along the $z$ and $y$ axes of the graphene plane.}
\label{fig:fig3}
\end{figure}

\begin{figure}
\includegraphics[scale=0.45]{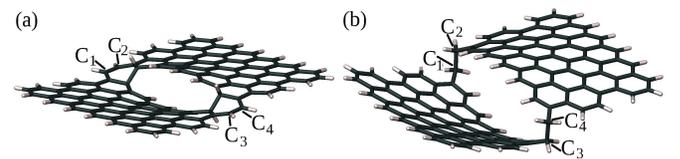}
\caption{ a) Unzipping of a flake when each carbon 
atom at the edges (including the C$_{1-4}$ atoms) is terminated by 
a single hydrogen atom. b) Unzipping of a flake for the case 
when C$_{1-4}$ carbon atoms are terminated by two hydrogen atoms before unzipping.}
\label{fig:fig4}
\end{figure}

For three or more pairs of the hydrogenated carbon bonds 
forming a line as presented in Fig.~\ref{fig:fig3} 
we found an interesting effect: The middle carbon-carbon bond from the 
line of hydrogenated atoms is being broken. 
To understand this behavior, we have to analyze the 
changes brought into the graphene lattice by hydrogenation.
If two carbon atoms connected by the carbon-carbon bond are hydrogenated, 
it alters the hybridization of this bond 
from $sp^2$ to $sp^3$, thereby enlarging its length from 1.42 \AA\
to 1.52 \AA\ \cite{bouk}. As a result, 
the strength of such a carbon-carbon bond is significantly lowered.
Moreover, hydrogenation causes the lattice distortion such 
that the carbon atoms hydrogenated from different sides of graphene 
are shifted out of its plane in opposite directions 
(in our present work it occurs along the $z$ - axis in Fig.~\ref{fig:fig3} (b)).
In the ideal honeycomb lattice the inter-atomic forces between the carbon atoms 
are almost compensated within the graphene plane 
(in this work the plane is formed in the $x$ and $y$ axes) 
and only the carbon atoms at the edges experience some
compressive stress \cite{jun}.
However, the shift of the hydrogenated carbon atoms out of the plane
leads to the appearance of the uncompensated inter-atomic forces 
arising on the modified bond and those forces are pointed out of the plane.
The forces function as the tensile stress applied 
to the $sp^3$ hybridized carbon-carbon bond (along the $z$ and $y$ axes)
thus additionally reducing its strength. 
When there are three pairs of hydrogen atoms attached to the 
carbon atoms as presented in Fig.~\ref{fig:fig3} (a),
the tensile stress arising at the modified carbon bonds is actually 
weaker for the first and last pairs
due to the compensation effect induced by the unmodified lattice around it. 
For the middle pair, the tensile stress is able to break the carbon-carbon bond. 

We calculated the inter-atomic forces 
for the ideal graphene lattice and found that 
they are zero in the $z$-direction, while 
in the $x$ and $y$-axes they are almost compensated within the graphene plane
(the remaining inter-atomic forces is $\sim\pm$ 0.009 eV/Bohr of magnitude).
Hydrogenation performed along the line as presented in Fig.~\ref{fig:fig3} (a)
in the $y$-direction increases the existing forces on the hydrogenated carbon atoms 
up to $\sim\pm$ 0.5 eV/Bohr, while in the $z$-direction
it causes an appearance of the inter-atomic forces 
of magnitude $\sim\pm$ 0.9 eV/Bohr (shown as $F_y$ and $F_z$ in Fig.~\ref{fig:fig3} (b)). 
Therefore, these forces in the $z$ and $y$-axes are larger
almost by two orders of magnitude than that in the ideal 
graphene lattice and the resulting tensile strain applied to the $sp^3$ hybridized bond 
(see $F$ in Fig.~\ref{fig:fig3} (b))
is $\sim\pm$ 1.0 eV/Bohr from each side. 
Such a tensile strain appears to be large enough to break 
the weak $sp^3$ hybridized carbon-carbon bond.

If we continue the line of 
hydrogenation presented in Fig.~\ref{fig:fig3} (a) to the edges, 
will the tensile stress disintegrate the flake?
We observed that the flake separation occurs 
only in the middle of the hydrogenation line 
(see Fig.~\ref{fig:fig4} (a)) but not at the armchair edges 
(see the C$_{1-4}$ carbon atoms in Fig.~\ref{fig:fig4}).
Away from the edges the carbon atoms are connected to
three nearest-neighbors and therefore, after their bonding 
with the hydrogen atom all their valence electrons are saturated thus converting
the carbon-carbon bond into the $sp^3$ hybridization type.
However, the carbon atoms at the armchair edges have only two 
neighbors and their bonding with one hydrogen atom turns the
C$_{1}$-C$_{2}$ and C$_{3}$-C$_{4}$ 
bonds into $sp^2$ hybridization type
(the bond length is 1.43 \AA) instead of the required $sp^3$ type. 
The termination of the 
C$_{1-4}$ carbon atoms by the two hydrogen atoms (see Fig.~\ref{fig:fig4} (b))
does not break the C$_{1}$-C$_{2}$ and C$_{3}$-C$_{4}$ bonds yet
but lead only to their elongation (the bond length is 1.58 \AA).
To break these bonds the termination of the
C$_{1-4}$ carbon atoms (dangling after bond separation)
by three hydrogen atoms is required
which can not be done within the graphene lattice preceding the 
bond breaking (within the graphene lattice 
those carbon atoms have only two valence electrons left). 

\begin{figure*}
\includegraphics[scale=0.90]{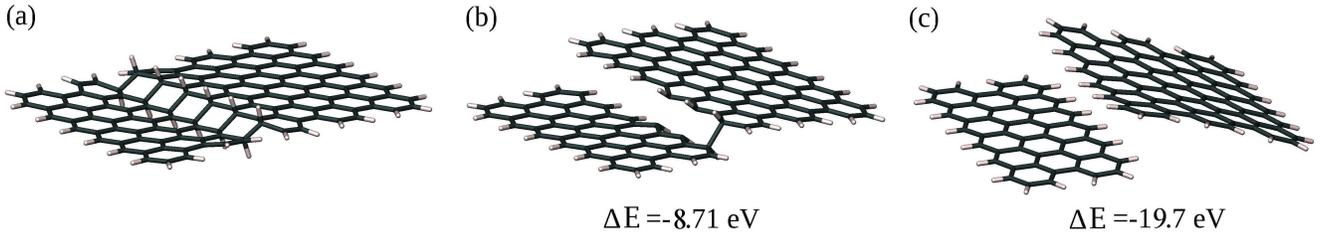}
\caption{The cutting of graphene flakes by formation of a
line of carbon atoms being hydrogenated from two sides of the graphene plane. 
a) Shift of the carbon atoms belonging to different 
sublattices of graphene in opposite direction out of plane 
and elongation of the carbon bond up to $\sim$1.75 \AA\ as a 
result of changing the hybridization of this bond from 
$sp^2$ to $sp^3$ type. b) Breaking of carbon-carbon bonds 
between hydrogenated carbon atoms as a result of
bond weakening and bond elongation turned on by the $sp^3$ hybridization
and origin of the tensile strain on this bond.
Shift of the flakes relative to each other. c) Migration of the 
flakes after their disintegration.}
\label{fig:fig5}
\end{figure*}

\begin{figure*}
\includegraphics[scale=0.80]{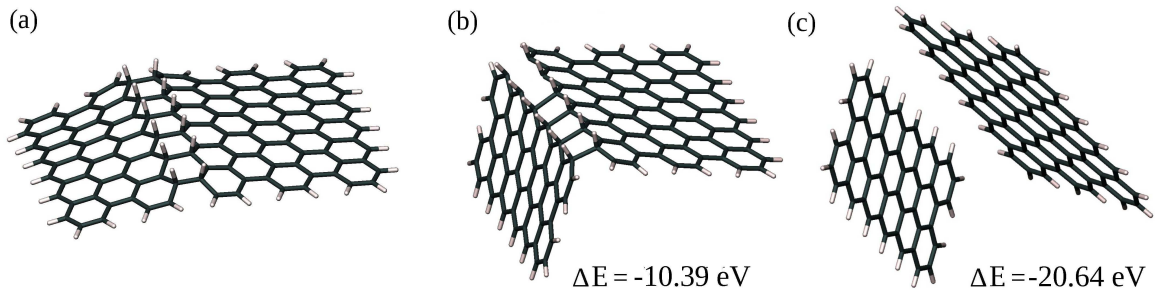}
\caption{The cutting of graphene flakes by formation of a
line of carbon-carbon bonds for which atoms are hydrogenated from one side of the graphene plane.
a) Bending of the lattice due to bonding of carbon atoms with the hydrogen atoms:
shift of all carbon atom positions in the graphene lattice occurs in the same direction,
and elongation of the carbon bond to $\sim$1.75 \AA\ as a 
result of changing of the carbon-carbon bond hybridization from 
$sp^2$ to $sp^3$ type. b) Breaking of the carbon-carbon bonds 
between hydrogenated carbon atoms due to the stress induced by 
modification of the lattice. c) Migration of the 
flakes after their disassembling.}
\label{fig:fig6}
\end{figure*}

However, if the line of hydrogenation is shifted as in 
Fig.~\ref{fig:fig5} (a), all the carbon atoms in the line are involved 
into bonding with three nearest-neighbors and attachment of the one extra
hydrogen atom would indeed turn all the bonds into the 
$sp^3$ hybridization type thereby 
causing their weakening and elongation. 
The tensile strain arising as a result of distortion 
of the graphene lattice breaks all the 
$sp^3$ bonds that would lead to disintegration of the original flake 
(see Fig.~\ref{fig:fig5} (b-c)). 
We calculated the evolution of the total energy ($\Delta E$) of the system 
as one flake is cut into two flakes. We obtained that 
its disintegration significantly lowers the total energy.
After the cutting is done (see Fig.~\ref{fig:fig5} (c)), the 
generated flakes take the position supporting the interaction 
of their dangling bonds left on the carbon atoms at the zigzag edges.

Technologically, it is easier to perform 
hydrogenation of the graphene flake from one side of the plane.
Taking into account the great advantage offered 
by graphene cutting through its hydrogenation,
we also investigated the behavior of the graphene system in the case of one-sided  
hydrogenation (see Fig.~\ref{fig:fig6}). 
We took the same flake as above and attached the hydrogen atoms at the same positions 
as in Fig.~\ref{fig:fig5} but all from the same side of the plane. 
In the case of one-sided hydrogenation,
the hydrogenated carbon atoms are shifted out of the plane but
in the same direction (see Fig.~\ref{fig:fig6} (a)). 
Similar to the double-sided hydrogenation,
one-sided hydrogenation alters the hybridization of the carbon-carbon bonds from 
the $sp^2$ to $sp^3$ type.
However, because the originated inter-atomic forces are all pointed in the same direction, 
they do not generate the tensile strain to the bonds but
initiate the bending of the flake along the 
line of $sp^3$ hybridized carbon-carbon bonds (see Fig.~\ref{fig:fig6} (b)). 
At some point the stress arising at the bend is large enough to break
these $sp^3$ carbon-carbon bonds (see Fig.~\ref{fig:fig6} (b)).
After the bonds are broken, similarly to previous double-sided 
case presented in Fig.~\ref{fig:fig5} (c),
the disintegrated flakes migrate into the position which maximizes the interaction
between the dangling bonds left on the carbon atoms at the zigzag edges 
that lowers the total energy of the system.

Therefore, for the planar graphene flakes 
if the carbon-carbon bonds 
are turned into the $sp^3$ hybridization type by hydrogenation, 
it significantly lowers the strength of these bonds.
Because hydrogenation induces lattice distortion thereby inhibiting the planarity of the flake,
it causes the occurrence of the stress which is large enough 
to break the $sp^3$ bonds if they form a line. 
If two sides of the graphene planes are involved in
hydrogenation, the resultant stress works as the tensile strain applied to the 
$sp^3$ bonds, while for one-sided hydrogenation, this stress initiates 
the flake bending that eventually breaks the bonds.
The proposed in-line hydrogenation of graphene can be successfully 
applied to cut the graphene flakes in two or more pieces
at the same time. Moreover, the same procedure can be used to unzip a
carbon nanotube into a flake characterized by the ideal zigzag edges that is
in agreement with previous investigations \cite{lu}.

\section{Conclusion}
We have shown that hydrogenation not only 
plays an important role 
in defining the electronic properties of graphene \cite{review},
it might also be used to control the graphene geometry. 
We found that flakes characterized by the linear and non-hydrogenated 
edges have the ability to merge into a bigger flake or a nanotube 
through zipping of their zigzag or armchair edges. 
Based on our results we suggest that 
a nanotube would be particularly characterized by the lower total energy because it
has an energetically favorable shape. 
The energetic preference of a tubular shape 
comes from the large number of carbon-carbon bonds to be made
within the zipping process
than that for planar geometries. 
Formation of each carbon-carbon bond 
lowers the total energy of the system by $\sim$ 6 eV.
This means that in case of high mobility of the 
carbon structures, to maintain the shape of others geometries 
such as graphene flakes,
hydrogenation has to be applied to all the edges.
However, hydrogenation of the graphene flakes has another tricky issue.
If its application on the graphene surface creates the line of 
carbon-carbon bonds whose both atoms are 
hydrogenated, it can lead to a separation of the flakes
into several pieces. The reason for such a behavior is that
hydrogenation causes the alteration of the bond hybridization 
from $sp^2$ to $sp^3$ type that significantly suppresses the bond strength 
and the resulting stress turned on by lattice distortion under some conditions can
be large enough to break those converted $sp^3$ bonds.
Such cutting of the graphene flakes can be performed through 
one-sided or double-sided hydrogenation of the graphene plane.

\section{Acknowledgments}
The work has been supported by the Canada Research Chairs Program of
the Government of Canada.

\end{document}